\documentclass[a4paper, conference]{IEEEtran}      

\IEEEoverridecommandlockouts                              

\usepackage{epsfig} 
\usepackage{amsmath} 
\usepackage{amssymb}  
\usepackage{txfonts} 
\usepackage{bm}

\newcommand{\argmin}{\operatornamewithlimits{argmin}}
\renewcommand{\vec}[1]{{\rm\bf{#1}}}

\begin{document}

\title{
Current Source Localization Using Deep Prior with Depth Weighting
\thanks{This work was supported in part by JSPS KAKENHI 21H05596.}
}

\author{\IEEEauthorblockN{Rio Yamana}
\IEEEauthorblockA{\textit{Graduate School of System Informatics} \\
\textit{Kobe University}\\
Kobe, Japan
}
\and
\IEEEauthorblockN{Hajime Yano}
\IEEEauthorblockA{\textit{Graduate School of System Informatics} \\
\textit{Kobe University}\\
Kobe, Japan
}
\and
\IEEEauthorblockN{Ryoichi Takashima}
\IEEEauthorblockA{\textit{Graduate School of System Informatics} \\
\textit{Kobe University}\\
Kobe, Japan
}
\and
\IEEEauthorblockN{Tetsuya Takiguchi}
\IEEEauthorblockA{\textit{Graduate School of System Informatics} \\
\textit{Kobe University}\\
Kobe, Japan
}
\and
\IEEEauthorblockN{Seiji Nakagawa}
\IEEEauthorblockA{\textit{Center for Frontier Medical Engineering} \\
\textit{Chiba University}\\
Chiba, Japan
}
}

\maketitle

\begin{abstract}
  This paper proposes a novel neuronal current source localization
  method based on Deep Prior that represents a more complicated prior
  distribution of current source using convolutional networks. 
  Deep Prior has been suggested as a means of an unsupervised learning
  approach that does not require learning using training data, and
  randomly-initialized neural networks are used to update a source
  location using a single observation.  In our previous work, a
  Deep-Prior-based current source localization method in the brain has
  been proposed but the performance was not almost the same as those
  of conventional approaches, such as sLORETA.  In order to improve the
  Deep-Prior-based approach, in this paper, a depth weight of the current
  source is introduced for Deep Prior, where depth weighting amounts
  to assigning more penalty to the superficial currents.  Its
  effectiveness is confirmed by experiments of current source
  estimation on simulated MEG data.

\end{abstract}




\section{INTRODUCTION}

Magnetoencephalography (MEG) and electroencephalography (EEG) are
non-invasive measurements of human brain activities that provide
excellent temporal resolution.  The estimation of current sources in
the brain using MEG and EEG has been helped to elucidate brain
function and assist in the diagnosis of brain diseases. However,
estimating the current distribution in the brain is inherently
difficult because it is an underdetermined problem with a small number
of MEG/EEG sensors relative to the number of current source
parameters.

Conventional methods for current source estimation, such as Minimum
Norm Estimation (MNE) \cite{MNE} and Standardized Low Resolution Brain
Electromagnetic Tomography (sLORETA) \cite{sLO}, solve this
problem by explicitly giving the prior distribution of the current
source. However, it is difficult to obtain the prior
distribution of the actual current sources, and estimation based on an
incorrect prior distribution may result in a large error.

In recent years, deep convolutional networks have been shown to play a
role in the prior distribution of natural images. One such unlearned
network is called ``Deep Image Prior'' and has been shown to be
effective for inverse problems in the image field \cite{DIP}.  In our
previous work~\cite{DBrainP}, we proposed a method for solving the
inverse problem of MEG and EEG using an untrained deep network prior
with a convolutional structure (Deep Prior), and showed that the
convolutional networks can represent the prior distribution of current
sources. However, the estimation error of the current source
localization using Deep Prior was not almost the same as that of the
conventional method, sLORETA, and there was still room for improvement
with Deep Prior.

In this paper, we propose a solution that takes into account the depth
weight of the current source to improve the current source estimation
accuracy using Deep Prior. It is known that the current density values
estimated by MNE tend to be higher near the brain surface, and the MNE
with depth weighting improves the localization accuracy.  Also, as the
solution obtained using Deep Prior may have some ``bias'', it
is expected to be improved by
taking the depth weight into consideration.  In this paper, we
evaluated the localization error of the current source
using artificially-generated MEG data assuming 
a single current source in the brain, and the performance of Deep Prior with
depth-weighted regularization is compared to those of coventional methods.

\begin{figure*}[t]
\begin{center}
  \includegraphics[scale=0.45]{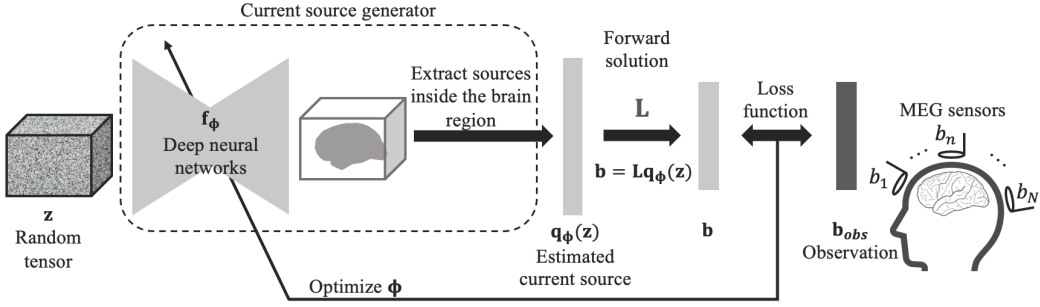}
\end{center}
\caption{Current source estimation in the brain using the deep prior.
  This method does not require learning using training data.  A
  randomly-initialized neural network is used to update a source
  location, and the input is a random tensor. The structure of neural networks is
  used as a current source prior.}
\label{fig:overview}
\end{figure*} 

\section{Formulation of Current Source Estimation}
\subsection{MEG Forward Problem}
Finding the magnetic field observed by sensors when current sources
in the brain is given is called a forward problem.  In this work, by
discretizing a given region in the brain and fixing the position of
the current source on the mesh point, the magnetic field ${\rm\bf{b}}
\in \mathbb{R}^{M}$ observed by the sensor can be expressed in the
form of the product of the lead field matrix ${\rm\bf{L}}$ and the
current vector ${\rm\bf{q}}\in \mathbb{R}^{3N}$:
\begin{eqnarray}
  {\rm\bf{b}} = {\rm\bf{L}}{\rm\bf{q}}
  \label{eq:order_prob}
\end{eqnarray}
where $M$ is the number of sensors and $N$ is the number of mesh points.
The lead field matrix is given by numerical calculation, such as the
boundary element method, which is based on the position of the sensor,
the position of the mesh point, and the conductivity in the brain.

\subsection{MEG Inverse Problem}

The inverse of the forward problem is to find the current source in
the brain from the observed magnetic field containing noise. This is
commonly referred to as the ``inverse problem''.  When the brain is
discretized, the number of current sources becomes very large compared
to the number of sensors. This makes it difficult to uniquely obtain
the current source from the observed magnetic field. This is also
called an ``ill-posed problem''.


Conventional methods such as MNE and sLORETA assume a
multi-dimensional normal distribution for the prior distribution of
noise and current sources contained in the observed values, and
minimize the sum of the error and the regularization term between the
forward problem and the observed value ${\rm\bf{b}}_{obs}$. It gives us a new
estimation $\hat{{\rm\bf{q}}}$:
\begin{eqnarray}
  \hat{{\rm\bf{q}}}
  & = & \argmin_{{\rm\bf{q}}} \left[ E_{\vec{C}}(\vec{Lq}; \vec{b}_{obs})
    + \lambda {\rm\bf{q}}^{\mathrm{T}}{\rm\bf{S}}^{-1}{\rm\bf{q}}\right]
  \label{eq:org_est_q}\\
  & = & \argmin_{{\rm\bf{q}}} \left[ ({\rm\bf{b}}_{obs} - {\rm\bf{L}}{\rm\bf{q}})^{\mathrm{T}}
      {\rm\bf{C}}^{-1} ({\rm\bf{b}}_{obs} - {\rm\bf{L}}{\rm\bf{q}}) \right. \nonumber\\
  &   & \left.  + \lambda {\rm\bf{q}}^{\mathrm{T}}{\rm\bf{S}}^{-1}{\rm\bf{q}} \right] \\
      & = & {\rm\bf{S}}{\rm\bf{L}}^{\mathrm{T}}({\rm\bf{L}}{\rm\bf{S}}{\rm\bf{L}}^{\mathrm{T}} + \lambda{\rm\bf{C}})^{-1} {\rm\bf{b}}_{obs}
\label{eq:est_q}
\end{eqnarray}
where ${\rm\bf{S}}$ is the covariance matrix of the parameters of the current
source, and ${\rm\bf{C}}$ is the covariance matrix of the noise in the
sensor. However, it is difficult to obtain the probability
distribution of the actual current source, and an estimation based on
a prior distribution that differs from the actual one may result in a
large error.

\section{Depth Weighting}

Since MNE minimizes the observation error under the L2 norm constraint
of the brain currents, the current density values estimated by MNE
tend to be higher near the brain surface.  In order to compensate for
it, a depth-dependent factor is introduced for the covariance matrix
${\rm\bf{S}}$ of the prior distribution of the currents \cite{D-Weight}.
\begin{equation}
s_{k}=\left(\vec{l}_{3k-2}^\mathrm{T}\vec{l}_{3k-2}
+\vec{l}_{3k-1}^\mathrm{T}\vec{l}_{3k-1}
+\vec{l}_{3k}^\mathrm{T}\vec{l}_{3k}\right)^{-p}
\label{eq:weight}
\end{equation}
where $\vec{l}_{i}$ is the $i$-th column vector of the lead field
matrix ${\rm\bf{L}}$ and $p$ is a depth weighting
parameter. $l_{3k-2}$, $l_{3k-1}$, and $l_{3k}$ correspond to the lead
fields of the $x$, $y$, and $z$ components of the $k$-th current
source, respectively. Since each component of the lead field matrix is
smaller for current sources located farther away from the sensor,
i.e., deeper in the brain, the corresponding value of $s_k$ becomes
larger.  Therefore, given by $s_k$, the
variance of the prior distribution of the currents at deeper locations
increases, making it easier to estimate the currents at deeper
locations.
Also, depth weighting amounts to assigning more penalty to the superficial
currents \cite{D-Weight}.

\section{Deep Prior with Depth Weighting}
When carrying out current source estimation using Deep Prior, the
current ${\rm\bf{q}}$ is generated by neural networks
$\vec{f}_{\bm\phiup}(\vec{z})$ with the latent variable $\vec{z}$ as input, and
the network parameters ${\bm\phiup}$ is estimated so that the observation
error is minimized. In our method, $\vec{q}$ in (\ref{eq:org_est_q}) is
replaced by the output $\vec{f}_{\bm\phiup}(\vec{z})$ of the neural networks and the
covariance matrix $\vec{S}$ of the current is modified using
(\ref{eq:weight}) in order to perform depth-weighted estimation. When
taking depth weight into consideration using Deep Prior to solve an
inverse problem, the solution of the current source estimation is as follows:
\begin{align}
  \hat{\bm\phiup} &= \argmin_{\bm\phiup}\left[E_{\vec{C}}(\vec{Lf}_{\bm\phiup}(\vec{z});\vec{b}_{obs})
    + \lambda \vec{f}_{\bm\phiup}(\vec{z})^\mathrm{T}\vec{S}^{-1}\vec{f}_{\bm\phiup}(\vec{z})\right] \label{eq:deep_prior}\\
  \hat{\vec{q}} &= \vec{f}_{\hat{\bm\phiup}}(\vec{z})
\end{align}
where each element of the latent variable $\vec{z}$ is sampled from
the standard normal distribution that is independent of each other.
The local optimum $\hat{\bm\phiup}$ is obtained by starting from a
random initialization of the parameter $\bm\phiup$. The only information
available is the observation vector from MEG sensors.

Fig.~\ref{fig:overview} shows the overview of the deep prior in this
work.  The number of the dimensions of the random input $\vec{z}$ was
set to 128.  The size of the final layer of the network corresponds to
the arrangement of mesh points. The number of channels in the final
layer was set to three corresponding to the $x$, $y$, and $z$
components of the current source. From the output of the final layer,
only the elements corresponding to the mesh points in the brain region
were extracted and used as the final output of the network (current
sources).

\section{Evaluation Experiment}

Current source estimation was performed on artificially-generated MEG
data. A head model of a subject and settings of a MEG system in
MNE-Python sample dataset \cite{MNE-Python} were used as the
simulation environment.  The MEG measurement system has a total of 306
sensors, consisting of 204 planar gradiometers and 102
magnetometers. A current source with a peak intensity of 50 nAm at 0.1
sec was placed in the center of the primary auditory and primary
visual cortex of the right hemisphere in the brain. (Current source
estimation was performed at 0.1 sec.)  The noise level was equivalent
to that after averaging over 80 MEG epochs. The noise covariance
matrix was computed from a noise recording in the dataset.
The peak signal-to-noise ratio was 21.6 dB.

MNE and sLORETA were implemented by MNE-Python~\cite{MNE-Python}.  The localization
error was defined as the Euclidean distance between the actual test
source and the estimated location of the maximum amplitude in the
estimated current source distribution.
The regions in the brain were discretized at 5 mm intervals,

\section{Results and Discussion}

\begin{figure}[tb]
 \begin{center} 
     \includegraphics[scale=0.4]{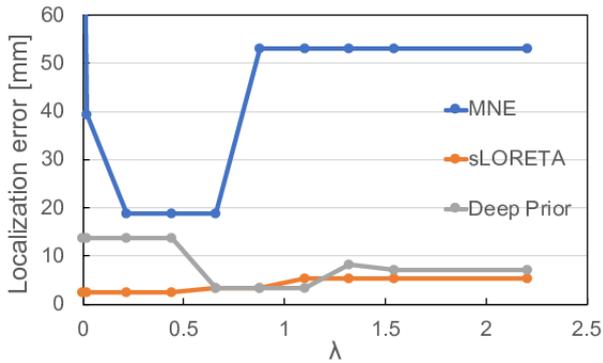}
  \caption{Relationships between the regularization parameter $\lambda$ and localization error of the current dipole in the right primary auditory cortex}
  \label{fig:lambda_aud}
 \end{center}
\end{figure}

\begin{table}[tb]
\caption{Localization error of the current dipole in the right primary auditory cortex}
\label{tab:loc_err_aud}
\begin{center}
\begin{tabular}{|c||c|c|}
\hline
Method     & \begin{tabular}{c} Localization\\ Error [mm]\end{tabular}\\
\hline\hline
MNE ($\lambda=0.44$) & 18.9 \\\hline
sLORETA ($\lambda=0.22$) & 2.5 \\\hline
Deep Prior ($\lambda=0$) & 13.7 \\\hline
Deep Prior ($\lambda=0.88$) & 3.4 \\\hline
\end{tabular}
\end{center}
\end{table}

Fig.~\ref{fig:lambda_aud} shows the localization error for the dipole
source in the right primary auditory cortex when varying the
regularization parameter $\lambda$.  Table~\ref{tab:loc_err_aud} shows
the localization error and the value of $\lambda$ for the smallest
error. Also, the localization error of Deep Prior for $\lambda = 0$
(no depth weight) is shown in Table~\ref{tab:loc_err_aud}.  As shown
in these results for the right primary auditory cortex, the
localization error of Deep Prior with depth weighting is almost the
same as that of sLORETA when setting an appropriate $\lambda$.

\begin{figure}[tb]
 \begin{center} 
     \includegraphics[scale=0.65]{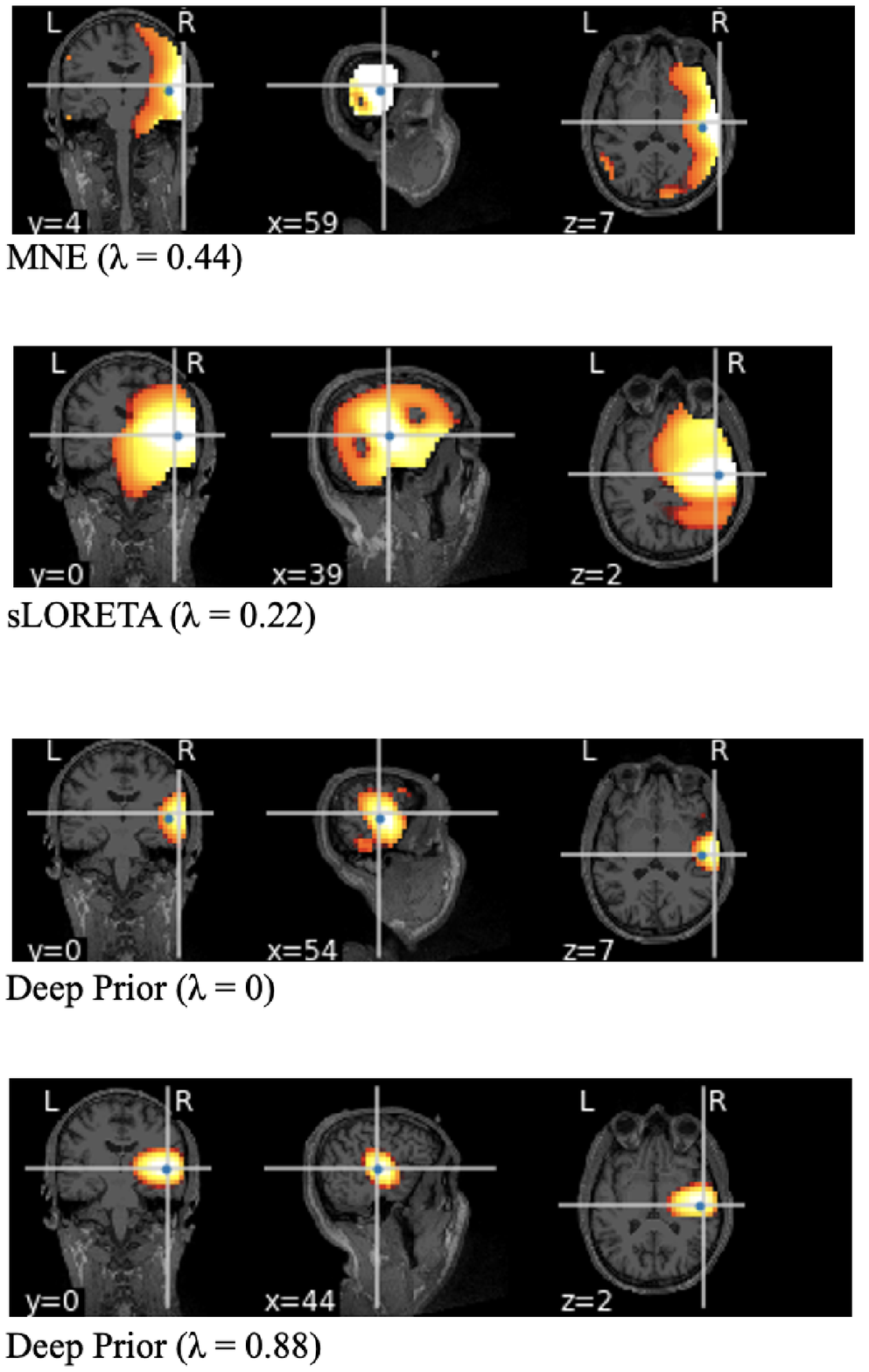}
  \caption{Estimated current source from MEG generated by the current dipole in the right primary auditory cortex}
  \label{fig:estimate2D}
 \end{center}
\end{figure}

The estimated current distribution is shown in
Fig.~\ref{fig:estimate2D}, where the actual current source is placed
in the right primary auditory cortex.  The three images in one box
show the current distribution estimated by each method in the $xz$,
$yz$, and $xy$ plane at $y$, $x$, and $z$ = ``the highest current
intensity,'' respectively. The blue dot in each image indicates the
coordinate of the actual current source.  As shown in
Fig.~\ref{fig:estimate2D}, when $\lambda = 0$, the current
distribution estimated by Deep Prior is distributed at shallow
locations in the brain.
On the other hand, the current
distribution estimated by Deep Prior with an appropriate $\lambda$ is
centered on the location of the actual current source.

\begin{figure}[tb]
 \begin{center} 
     \includegraphics[scale=0.4]{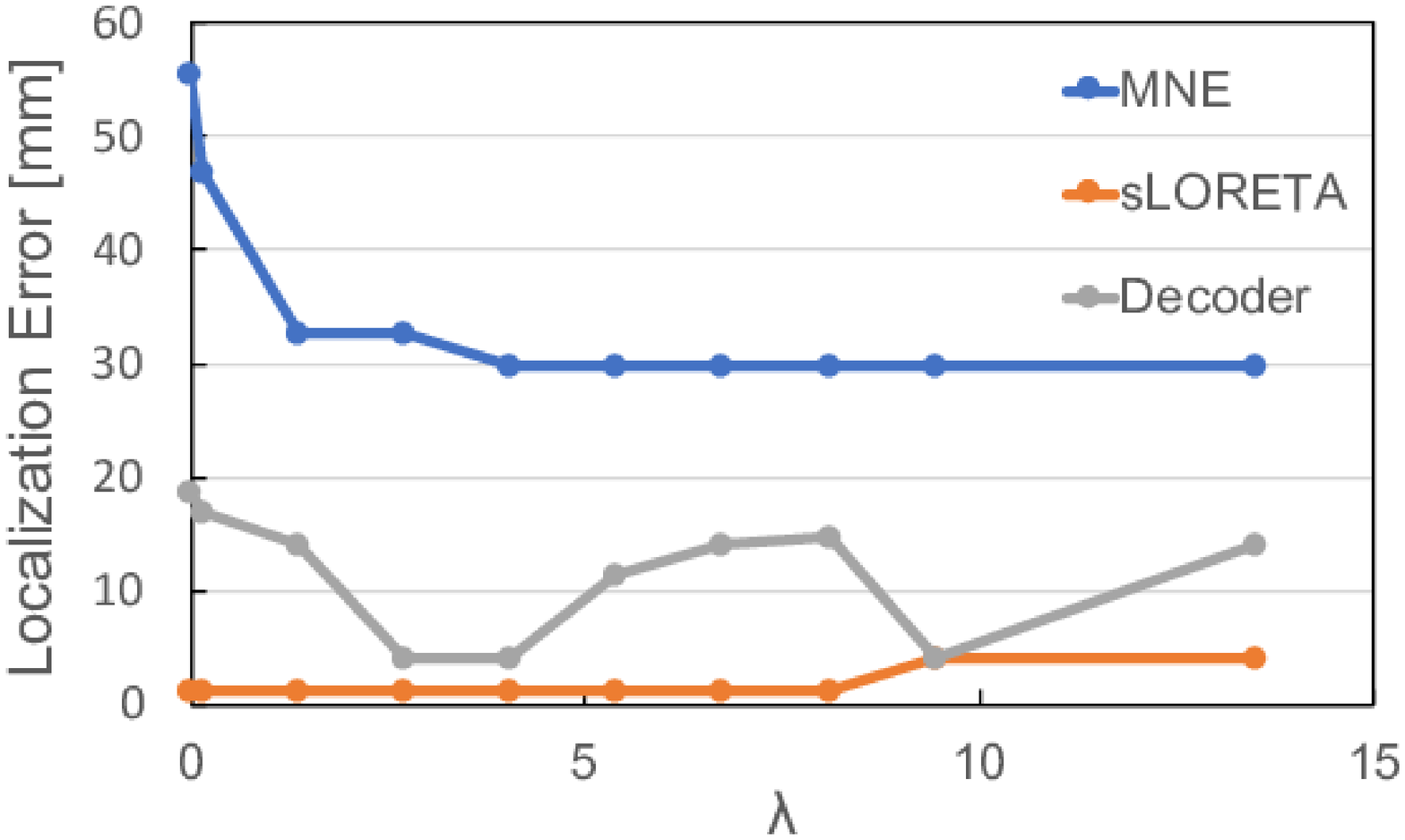}
  \caption{Relationships between the regularization parameter $\lambda$ and localization error of the current dipole in the right primary visual cortex}
  \label{fig:lambda_vis}
 \end{center}
\end{figure}

\begin{table}[tb]
\caption{Localization error of the current dipole in the right primary visual cortex}
\label{tab:loc_err_vis}
\begin{center}
\begin{tabular}{|c||c|c|}
\hline
Method     & \begin{tabular}{c} Localization\\ Error [mm]\end{tabular}\\
\hline\hline
MNE ($\lambda=4.05$) & 29.7 \\\hline
sLORETA ($\lambda=4.05$) & 1.8 \\\hline
Deep Prior ($\lambda=0$) & 18.6 \\\hline
Deep Prior ($\lambda=4.05$) & 4.2 \\\hline
\end{tabular}
\end{center}
\end{table}

Fig.~\ref{fig:lambda_vis} shows the localization error of the
estimated current source when varying the regularization parameter
$\lambda$, where the actual current source is placed in the right
primary visual cortex. Table~\ref{tab:loc_err_vis} shows the
localization error and the value of $\lambda$ when the smallest
localization error was obtained for each method and $\lambda$ $= 0$
for Deep Prior.  As shown in Table~\ref{tab:loc_err_vis}, in that case
of the right primary auditory cortex as well, the localization errors
of the MNE and the Deep Prior reduced compared to the
case when $\lambda = 0$ by setting an appropriate $\lambda$.
However, the localization errors of MNE and Deep
Prior tended to be larger than that of the right primary auditory
cortex.  This may be due to the fact that the primary visual cortex is
located at a deeper position than the primary auditory cortex, and
increasing $\lambda$ also increases the regularization of the current
norm, i.e., the bias toward the surface. Therefore, it is necessary to
adjust not only $\lambda$ but also $p$ in (\ref{eq:weight}).

\section{CONCLUSIONS}

In this work, in order to improve the performance of current source estimation
using Deep Prior, we introduced regularization with depth weight in
Deep Prior. In experiments, the MEG data synthesized by assuming a
single current source in the right primary auditory cortex and the
right primary visual cortex are used, and the results showed that by
setting appropriate regularization parameters, the localization error
reduced and the current source could be estimated around the true
position.  Since the estimation is not yet satisfactory for deeper
current sources, it is necessary to investigate the effect of other
parameters $p$ and how to select optimal parameters. The evaluation of
the proposed method on a wider variety of current sources is also an
issue for future work.

\addtolength{\textheight}{-12cm}   



\begin{thebibliography}{99}

\bibitem{MNE} M. S. H{\"a}m{\"a}l{\"a}inen and R. Ilmoniemi,
  ``Interpreting measured magnetic fields of the brain: Estimates of
  current distributions,'' Technical Report TKK-F-A559 HUT Finland, vol.~32, 1984.
\bibitem{sLO} R. D. Pascual-Marqui,
  ``Standardized low-resolution brain electromagnetic tomography
  ({sLORETA}): technical details,'' Methods and findings in
  experimental and clinical pharmacology, vol.~24 Suppl D, pp.~5--12, 2002.
\bibitem{DIP} D. Ulyanov, A. Vedaldi and V. Lempitsky, 
  ``Deep image prior,''
  Int. J. Comput. Vis. vol.~128, pp.~1867--1888, 2020.

\bibitem{DBrainP}
  R. Yamana, H. Yano, R. Takashima, T. Takiguchi, and S. Nakagawa,
  ``MEG source localization Using Deep Prior,'' IEEE LifeTech, 2022.
\bibitem{D-Weight}F.-H. Lin, T. Witzel, S. P. Ahlfors,
  S. M. Stufflebeam, J. W. Belliveau, and
  M. S. H{\"a}m{\"a}l{\"a}inen, ``Assessing and improving the spatial
  accuracy in MEG source localization by depth-weighted minimum-norm
  estimates,'' NeuroImage, 31, pp.~160–171, 2006.

\bibitem{MNE-Python} A. Gramfort, M. Luessi, E. Larson,
  D. A. Engemann, D. Strohmeier, C. Brodbeck, R. Goj, M. Jas,
  T. Brooks, L. Parkkonen, and M. S. H{\"a}m{\"a}l{\"a}inen, ``MEG and
  EEG data analysis with MNE-Python,''
  Frontiers in Neuroscience, vol.~7, no.~267, pp.~1-13, 2013.

\end{thebibliography}
\end{document}